\documentclass[journal]{IEEEtran}
\usepackage[pdftex]{graphicx}
\usepackage{subcaption}
\ifCLASSINFOpdf
\else
\fi
\usepackage{amsmath}
\usepackage{xcolor,soul}
\usepackage{paralist}
\usepackage{amsmath,amssymb,amsthm}
\usepackage{makecell}

\definecolor{orange}{RGB}{230,97,1}
\definecolor{light orange}{RGB}{253,184,99}
\definecolor{lavender}{RGB}{178,171,210}
\definecolor{purple}{RGB}{94,60,153}

\newcommand{\hlc}[1]{{\sethlcolor{cyan!0}\hl{#1}}}

\begin{document}
\title{Did you miss it? Automatic lung nodule detection combined with gaze information improves radiologists' screening performance}

\author{Guilherme Aresta,
        Carlos Ferreira,
        Jo\~{a}o Pedrosa,
        Teresa Ara\'{u}jo,
        Jo\~{a}o Rebelo,
        Eduardo Negr\~{a}o,
        Margarida Morgado,
        Filipe Alves,
        Ant\'{o}nio Cunha,
        Isabel Ramos and 
        Aur\'{e}lio Campilho
        % <-this % stops a space
\thanks{G. Aresta is with the Faculty of Engineering of University of Porto (FEUP) and with the Institute for Systems and Computer Engineering, Technology and Science (INESC TEC) and is funded by the FCT grant contract SFRH/BD/120435/2016. guilherme.m.aresta@inesctec.pt.}% <-this % stops a space
\thanks{C. Ferreira and J. Pedrosa are with INESC TEC; T. Ara\'{u}jo is with FEUP and INESC TEC and is funded by the FCT grant contract SFRH/BD/122365/2016;  J. Rebelo, E. Negr\~{a}o, M. Morgado, Filipe Alves and Isabel Ramos are with the Faculty of Medicine of University of Porto (FMUP). A. Cunha is with INESC TEC and University of Trás-os-Montes e Alto Douro (UTAD). A. Campilho is with INESC TEC and FEUP.}% <-this % stops a space
\thanks{This study is associated with LNDetector, which is financed by the ERDF - European Regional Development Fund through the Operational Programme for Competitiveness - COMPETE 2020 Programme and by the National Fundus through the Portuguese funding agency, FCT - Funda\c{c}\~{a}o para a Ci\^{e}ncia e Tecnologia within project POCI-01-0145-FEDER-016673.}
}

\maketitle

\begin{abstract}
Early diagnosis of lung cancer via computed tomography can significantly reduce the morbidity and mortality rates associated with the pathology. However, search lung nodules is a high complexity task, which affects the success of screening programs. Whilst computer-aided detection systems can be used as second observers, they may bias radiologists and introduce significant time overheads. With this in mind, this study assesses the potential of using gaze information for integrating automatic detection systems in the clinical practice. For that purpose, 4 radiologists were asked to annotate 20 scans from a public dataset while being monitored by an eye tracker device and an automatic lung nodule detection system was developed. Our results show that radiologists follow a similar search routine and tend to have lower fixation periods in regions where finding errors occur. The overall detection sensitivity of the specialists was $\mathbf{0.67\pm0.07}$, whereas the system achieved 0.69. Combining the annotations of one radiologist with the automatic system significantly improves the detection performance to similar levels of two annotators. Likewise, combining the findings of radiologist with the detection algorithm only for low fixation regions still significantly improves the detection sensitivity without increasing the number of false-positives. The combination of the automatic system with the gaze information allows to mitigate possible errors of the radiologist without some of the issues usually associated with automatic detection systems.

\end{abstract}

\begin{IEEEkeywords}
Lung cancer, computer-aided diagnosis, eye-tracking, deep learning, clinical environment
\end{IEEEkeywords}

% \fi
\IEEEpeerreviewmaketitle

\section{Introduction}

%checked
Lung cancer is the deadliest type of cancer worldwide, but the morbidity and mortality rates can be significantly reduced if the diagnosis is performed early enough. Namely, screening programs with chest low-dose computed-tomography (CT) images of risk-groups have shown to reduce mortality more than 20\% in relation to chest radiography~\cite{Screening2015}. 
During the screening process, trained radiologists search for pulmonary nodules, primary indicators of lung cancer, inside the lung parenchyma. 

%checked
Lung cancer screening is non-trivial because lung nodules can present a wide range of opacities (commonly referred as textures), shapes, dimensions and locations, and thus the experience of the specialist tends to play an important role on the success of the nodule hunting and corresponding characterization~\cite{MacMahon2017}. Furthermore, CT scans are inherently complex to analyze due to their 3-dimensionality and large range of intensities to explore, making the process tiresome and consequently more prone to errors.

Radiologists fail at nodule detection either due to fixation or recognition errors~\cite{Krupinski2010CurrentPerception}. Fixation errors, mostly related to stress and fatigue, occur when the expert does not focus a region-of-interest for an enough period of time to identify potential nodule candidates.
On the other hand, recognition errors result from failing to correctly identify a found abnormality as a nodule and depends mostly on experience of the radiologist~\cite{Brunye2019Eye-trackingCardiologists}.

Assessing the gaze of radiologists during the screening process provides important information on why failures occur, and thus may be used for improving the overall success of the procedure.
Namely, eye-tracking equipments allow to record the spatial position of the radiologist's gaze during the analysis of the scan analysis, providing insight on how screening is performed.
For instance, it is known that radiologist usually follow on of two distinct nodule search strategies: scanning and drilling. In scanning, a radiologist searches for nodules on an entire slice before moving to the next, thus having to recur to techniques as maximum intensity projection (MIP) to assess depth information. Alternatively, in drilling the radiologist focus on a single quadrant of the volume at the time, scrolling through all the slices of the scan to account for 3D information~\cite{Drew20003, Diaz2015Eye-trackingData}.

Lung cancer computer-aided detection and diagnosis (CADe and CADx) systems can help to further increase the success of screening programs by identifying potential abnormalities to the radiologists and mitigating fixation-related failures. Also, the demand for these CADe systems has been raising due to the increase on the number of patients and the consequent equipment and trained personnel costs. CADe systems operate by automatically identifying potential nodules in the CT scan, which are then assessed by the radiologist. Because of this, a high detection sensitivity and low false-positive rates are essential characteristics of these systems. Given the complexity of the task, deep learning-based approaches are becoming the backbone of lung CADe systems since they allow to significantly reduce the field knowledge required to design efficient solutions.
Lung cancer CADe systems are usually composed of two stages: 
\begin{inparaenum}
\item a high sensitivity/low specificity 3D or 2D object detection framework, such as Faster-R CNN~\cite{Ren2017}, that guarantees the detection of the majority of the nodules, at the cost of also detecting other structures such as blood vessels or scars, and
\item a false-positive reduction neural network to remove the non-nodules proposed by the nodule detector.
\end{inparaenum}
A properly trained system allows to achieve detection sensitivities greater than $0.80$ with $0.125$ FP/scan or greater than $0.90$ with 1~FP/scan~\cite{Ding2017}.

%https://www.ncbi.nlm.nih.gov/pubmed/23300205
Despite the high detection performance of lung nodule CADe systems, their success as stand-alone tools in clinical practice is limited. \hlc{Indeed, human supervision can ensure the relevance of the findings, allowing to re-plan or even avoid unnecessary follow-ups.} Also, CADe systems tend to fail on cases that significantly deviate from the training data, namely unseen types of abnormalities.
Because of this, CADe systems are used by radiologists either as an indicator of regions-of-interest or as a second independent observer.

When used collaboratively, CADe systems can bias the decision process of the radiologist. Namely, checking a case for the first time with the CADe markings on it can lead the expert to focus their attention on the highlighted regions in detriment of the remaining scan. Furthermore, less experienced experts may over-trust the proposals of the CADe and increase the number of false-positive detections. On the other hand, \textit{a posteriori} review of CADe suggestions may introduce a large time overhead. 
\hlc{In this scenario, adjusting CADe results according to the attention and experience of specialists is of interest since it allows to mitigate CADes' drawbacks without compromising the success of the screening routine.} Namely, the integration of eye-tracking information with \hlc{CADe has been proposed, showing} promising results. Specifically, recent studies have shown that the gaze of the radiologists during the nodule \hlc{search} task can be used for establishing a set of nodule candidates, which can then be classified by a deep learning system as nodule/non-nodule with state-of-the-art performance~\cite{Khosravan2019ALearning}. 

%\begin{figure}
%    \centering
%    \includegraphics[width=1\columnwidth]{figures/todo.pdf}
%    \caption{TODO: overview of the experimental set-up}
%    \label{fig:my_label}
%\end{figure}

This study assesses the performance of 4 young radiologists on the lung nodule hunting task and how a CADe system can contribute to improve their success. For that purpose, gaze information recorded via an eye-tracker on a clinical setting is used for understanding how the search is conducted and how the radiologists' experience affects the process. Likewise, inter-observer evaluations are also conducted. Finally, it is shown that using a deep learning-based CADe system as a posterior second observer, both independently and together with gaze information, allows to improve the global nodule detection sensitivity without increasing the number of false-positives. 
The experimental setup, including the acquisition setting, gaze processing and the lung nodule detection algorithm is described in Section~\ref{sec:methods}. Section~\ref{sec:results} details the results of the reading sessions and the impact of the CADe system on the detection sensitivity. Finally, Section~\ref{sec:discussion} discusses and Section~\ref{sec:conclusion} summarizes the findings of this study.

\section{Materials and methods}
\label{sec:methods}

\subsection{Annotation procedure}

The annotation team is composed of 4 radiology interns from Hospital de S\~{a}o Jo\~{a}o, Porto, Portugal - Rad1, Rad2, Rad3, Rad4 - with experience between 1 and 4 years.
%- with 2, 3, 1 and 1 years of experience, respectively.
Each medical expert was asked to annotate the scans similarly to the first step of the LIDC-IDRI annotation protocol~\cite{Armato2011}.
Namely, the radiologists were instructed to mark every non-nodule and nodule with diameter $\leq 3mm$ with a point on the abnormality's center of mass and segment voxel-wise all lung nodules with diameter $\geq 3mm$. \hlc{For each of the abnormalities}, the radiologists were also asked to perform a subjective categorical characterization of the nodules' calcification pattern and internal structure (soft tissue, fluid, fat, air) and ordinal characterization of how well defined the margin is, the extent of the spiculation, their sphericity and lobulation, expected malignancy, \hlc{subtlety} and their texture (solid, sub-solid or non-solid)~\cite{McNittGray20071464}. For this purpose, a custom version of the ITK-SNAP software~\cite{py06nimg} was used. This custom version allows to retrieve, at a fixed sample rate, the physical pixel size of the view windows (axial, coronal and sagittal), the scan-wise coordinates of the slices currently under analysis as well as the respective pan and zoom settings. The annotation procedure was blind, \textit{i.e.} the radiologists did not have access to the ground-truth and could not discuss their markings with the other annotators.

\begin{figure*}
    \centering
    \includegraphics[width=\textwidth]{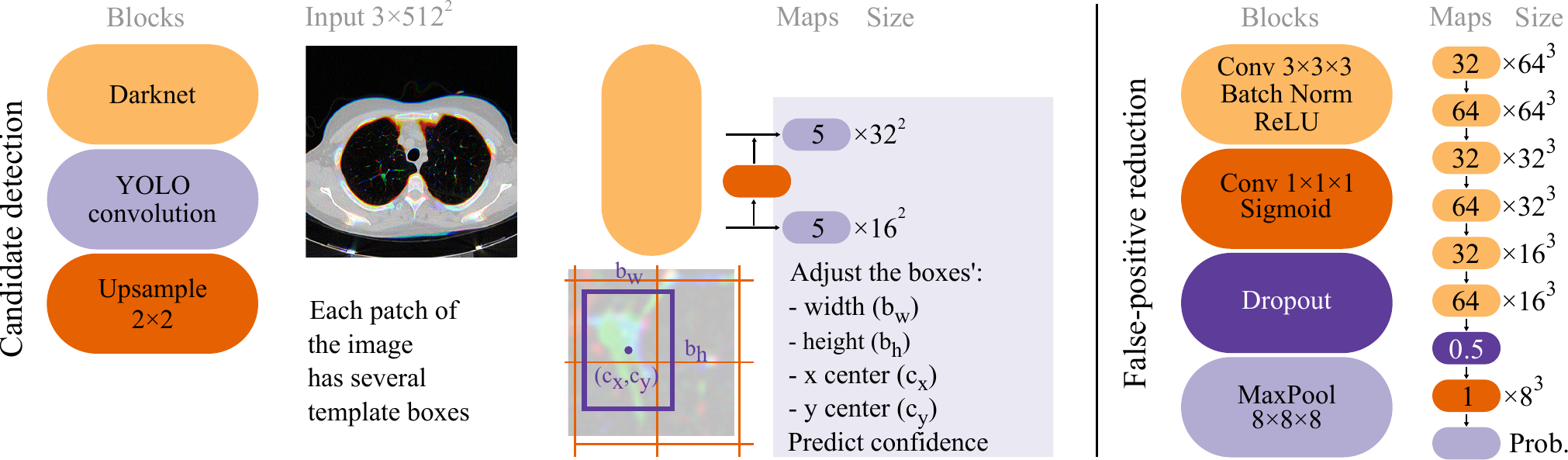}
    \caption{Schematic representation of the automatic lung nodule detection system's architecture, composed of an initial candidate detection followed by a false-positive reduction step.}
    \label{fig:model}
\end{figure*}

\subsection{Gaze capturing and processing}

The gaze of the radiologists was recorded using a Tobii Eye Tracker 4C (frequency $f$=90Hz) attached at the base of a Fujitsu E22T-7 monitor (1920$\times$1080 pixels). The sensor records the absolute position in physical screen pixels of the gaze.
%The radiologists were asked to always seat at the same distance from the monitor (head to monitor distance of 60 cm) and calibrate the sensor prior to each annotation session. No other requirements were made as to better simulate a truthful CT screening routine. Likewise, all annotations were performed in a radiology room (\textit{i.e.} with reduced lightning and limited access to avoid distractions). The radiologists used mainly the axial view to search for the lung nodules, using the other views for confirmation purposes.
The radiologists were asked to seat at a distance of 60 cm from the monitor in a room with reduced lightning and limited access to avoid distractions, and calibrate the sensor prior to each annotation session. Since the axial view was the one mainly used for nodule hunting, segmentation and characterization, all gaze points outside the window containing this view and those corresponding to the annotation procedure were \hlc{removed}. \hlc{On this window, all gaze points outside the lung volume were also removed. The lung volume mask was estimated by performing a fixed Hounsfield Unit threshold followed by a morphological closing operation to fill the gaps created by nodules and, blood vessels and other structures.}

Let $X,Y,Z$ be the dimensions, in voxels, of an analyzed scan. The coordinates of the gaze for each time point $t_i \in t$ are converted to integer scan-wise voxel coordinates $(x_i,y_i,z_i) \in G(t)$ having in account the current slice of the observation, $z$, as well as the respective zoom and pan. The corresponding attention map for each slice $z$, $\mathcal{A}_z$, is defined as in Eq.~\ref{eq:attention_z}:

\begin{equation}
    \mathcal{A}_z= \sum_{\sigma_k \in \boldsymbol{\sigma}} \left[ \left[ \sum_{j=1}^{|G_z|} \left[\left(\mathbf{O}_{G_{z_j}}\right)_{j-1}+1\right] \right] * \mathcal{N}(0,\sigma_k^2I) \right]
    \label{eq:attention_z}
\end{equation}

%window size = 3 standard deviations of the mean
%foveal vision 5.2 degree
%https://en.wikipedia.org/wiki/Peripheral_vision
\noindent where $I$ \hlc{is the identity matrix}, $\mathbf{O}$ is a $X \times Y$ zero-valued matrix, $G_z = \{\left(y_i,x_i\right): z_i=z\} \subset G(t)$, and $\sigma~\text{(voxel)} \in \boldsymbol{\sigma} \propto$~zoom is the spread of the isotropic multivariate normal distribution $\mathcal{N}$, $\sum \mathcal{N}(0,\sigma^2I) = 1/f$. Assuming that the radiologists' gaze is approximately the foveal vision (approximately 5$^\circ$~\cite{millodot2014dictionary}), \hlc{the values of $\boldsymbol{\sigma}$ are computed so that $99.7\%$ of $\mathcal{N}$ is contained inside a circle of diameter $5.2$~cm, \textit{i.e.} $6\sigma\implies5.2$~(cm), $\forall \sigma \in \boldsymbol{\sigma}$. This way, the voxel-wise diameter of $\mathcal{N}$ corresponds to the expected physical dimension of the gaze, $d$, when considering the head to monitor distance and the foveal vision angle}. Finally, the attention map $\mathcal{A}$ is defined as the concatenation of all attention slices:
\begin{equation}
    \mathcal{A} = \Vert_{z=1}^Z \mathcal{A}_z=\mathcal{A}_1\Vert \ldots \Vert \mathcal{A}_Z
\end{equation}

\noindent \hlc{where $\Vert$ is the concatenation operator}. $\mathcal{A}$ is a $X \times Y \times Z$ matrix where each element indicates an estimate of the duration, in seconds, of the observation of the respective voxel in the scan.

\subsection{Datasets}
\label{sec:datasets}

The performance of the annotators was evaluated on the LIDC-IDRI dataset~\cite{Armato2011}. These scans have been assessed by 4 radiologists, first blindly and afterwards with the markings of their peers. 
The subset of LIDC-IDRI considered for this study follows the LUNA16 Challenge~\cite{Setio2016}. Namely, an annotation was considered to be a nodule if at least 3 medical experts agreed on the diagnosis. The remaining lesions were considered as non-nodules. All nodules were subjectively characterized by each specialist from \hlc{1 to 6 in terms of calcification and 1 to 5 in terms of} internal structure, lobulation, expected malignancy, margin, sphericity, spiculation, subtlety and texture (non-solid, sub-solid and solid). In total, this study considers 20 scans with 42 nodules with radius $\geq 1.5mm$ of known center-of-mass and equivalent radius. %(see Fig.~\ref{fig:all_nodules}).
Also, the scans have an average number of slices of $299\pm142$, slice thickness of $1.29\pm0.62$(mm) and axial resolution of $0.67\pm0.07$(mm/voxel).

Besides the 888 scans from the LUNA Challenge, \hlc{294 thin-slice scans from proprietary dataset were also used for developing the automatic detection method}. All images were adquired by several Siemens models at Centro Hospitalar de S\~{a}o Jo\~{a}o. All scans have voxel-wise annotations (single blind) and most of the volumes where assessed by 2 of the radiologists that participated in this study. The total number of annotated nodules used was 985. The scans have an average number of $321\pm41$ slices, slice thickness of $1.00\pm0.08$~mm and axial resolution of $0.63\pm0.09$~mm/voxel.

\subsection{Deep CNN for automatic lung nodule detection}

The studied nodule detection system is composed of an initial candidate detector followed by a false-positive reduction step, as shown in Fig.~\ref{fig:model}.
The detection algorithm is based on the YOLOv3 architecture~\cite{Redmon2018YOLOv3:Improvement,10.1007/978-3-030-00946-5_31} and outputs bounding boxes of potential nodules on the scan. The model assumes that each patch of the input image can have at least one object of interest. Instead of predicting the bounding boxes from scratch, the nodule detection is performed by adjusting the dimensions and positions of several template boxes assigned to the same patch. Furthermore, given the wide range of the nodules' diameters, the prediction is performed by assessing feature maps of the network at 2 different scales in a pyramidal fashion, \textit{i.e.} at each scale the object location prediction results from the processing of the current set of feature maps as well as the ones from the previous scales. This increases the robustness of the model to variations in the size of the nodules. At each scale, the features maps are convolved to a $G\times G \times (T \times 5$ tensor), where $G\times G$ is the number of patches (function of the model's architecture), $T$ is the number of template bounding boxes and 5 is the number of parameters to optimize (the horizontal and vertical displacement of the bounding box, its width and height and the confidence of containing a nodule).
The detection network is trained only on slices containing nodules by minimizing a detection loss $\mathcal{L}_{\,\text{det}}$ \hlc{for each of the scales}:

\begin{equation}
    \mathcal{L}_{\,\text{det}} =
    \sum_{s=1}^2 \lambda_s \left(
    \alpha_1\mathcal{M}_{\text{\,centers}}+\alpha_2\mathcal{M}_{\text{\,dimensions}}+\alpha_3\mathcal{M}_{\text{\,confidence}} \right)
\end{equation}

\noindent where $\mathcal{M}$ is the mean square error, \hlc{$\mathcal{M_\text{\,centers}}$, $\mathcal{M_\text{\,centers}}$ and $\mathcal{M_\text{\,confidence}}$ are the loss components associate with the centroid, width/height and nodule presence of the bounding box, respectively,} and $\alpha_i$ are predefined weights.

The input to the model is a $512\times512\times3$ image composed of 3 neighbour axial slices~\cite{10.1007/978-3-030-00946-5_31} of the CT scan to reduce the complexity of the model and take advantage of transfer learning approaches. Specifically, lung nodule hunting is non-trivial due to the large amount of information to process and the existence of blood vessels, which circular cross-sections in the axial slices may act as nodule confounders. A possible solution is to use 3D networks that, by assessing the data in a volumetric fashion, ease the distinction of spherical nodules from the cylindrical blood vessels. However, 3D approaches are computationally heavy, hindering their application in clinical settings without recurring to cloud-based solutions. Also, it is known that fine-tuning pre-trained networks on natural images for medical image problems eases the training process. By using 3 neighbour axial slices, the system can take advantage of pre-trained networks for feature extraction and still encode depth information, as depicted in Fig.~\ref{fig:model}. For thin-slice CT scans, increasing the distance between neighbour slices allows to simulate higher slice thickness, helping to both increase the 3D context and standardize the training data.

The optimizer is Adam~\cite{DBLP:journals/corr/KingmaB14} and the data is augmented by random crops, rotations, translations and small color alterations (so that the 3D information is not lost). 
Also, hard samples mining is performed by, at the end of each epoch, increasing the probability of assessing images with higher prediction error on the previous iteration.
%To improve the robustness of the model, hard samples mining is performed. Specifically, at the end of each epoch all images of the training set are assessed. Images with higher prediction error are given a higher probability of belonging to a batch and are thus more likely to be seen by the network. \hl{reduzir}

After training, scan-wise inference is performed by sliding through all the slices of the volume. 
The network predicts, for each slice, candidates characterized by their bounding boxes and the respective probability of containing lung nodules. 
To reduce the number of false-positive detections, all candidates outside the lung volume, computed via a Hounsfield units-based threshold followed by a morphological closing, are discarded. 
The remaining predictions for which inter-centroid distance is less than half the size of their bounding box are merged by averaging their centroid and maintaining the highest network object \hlc{detection} probability.

The false-positive reduction network, summarized in Fig.~\ref{fig:model}, is trained based on the results of the nodule detection algorithm.
Namely, the training dataset is composed of all the nodules from the ground-truth and \hlc{$5\times$} the highest score false-positives of each scan. The input to the network is a cube of $51\times 51\times 51 (mm)$, resized to $64\times 64 \times 64$~(voxels), centered on the candidate's centroid. 
%The model has in-built explainability to improve its applicability in the clinical practice. 
The model considers the binary non-nodule/nodule classification as a multiple-instance learning (MIL) problem, which leads to a weight optimization via the minimization of the loss function:

\begin{equation}
    \mathcal{L}_{\,\text{fp}} = -\frac{1}{N}\sum_{n=1}^N \left[ y_n \text{log}\left(\texttt{m}\left(\mathbf{P}\right)\right) + \left(1-y_n\right)\text{log}\left(1-\texttt{m}\left(\mathbf{P}\right)\right) \right]
\end{equation}

\noindent where $y_n$ is the binary label of image $n$ (non-nodule or nodule), $N$ is the number of images, $\texttt{m}$ is the global max pooling operation and $\textbf{P}\in[0\,,1]$ is the last layer of the false-positive reduction network. Adam is used as optimizer and the dataset is artificially augmented via random crops, translations, flips and rotations.

The initial candidate detection was trained on the proprietary dataset (Section~\ref{sec:datasets}). This was done to avoid a potential overfit to the annotation style of LIDC-IDRI dataset, which could result on an over-estimation of the system's performance. \hlc{On the other hand, preliminary results of the false-positive reduction network trained with the proprietary dataset suggested that the model was not generalizing well to changes on the reconstruction kernels and new slice thickness observed in the test set. A poor generalization ability on different equipments was also present. Indeed, this is a known problem of deep learning systems, but fine-tuning on independent samples acquired with equipments where the model will be tested helps to mitigate the issue}~\cite{DeFauw2018ClinicallyDisease}. \hlc{With this in mind, this second network was trained on samples from the LIDC-IDRI dataset.}

\subsection{Performance evaluation}

\hlc{This study focuses on the nodule search technique of the radiologists during lung cancer screening, as well as the nodule detection performance without and with the automatic detection system. The details of the evaluation procedure are detailed on the next paragraphs.}

\subsubsection{Search technique} \hlc{The search technique is qualitatively and quantitatively evaluated by assessing the position of the gaze on the left and right lungs. For that, the left and right lungs were estimated by dividing the scan on sagital direction in the location corresponding to the mean of the minimum and maximum transverse points of the segmentation mask. Also, for each scan, the search time of each point, $t_i\in t$ was normalized to $t'_i\in[0\,,1] = t_i/T$, where $T$ is the the total scan reading time. Then, for all scans, the normalized time points were sampled to 100 points. The probability of the gaze being located on the right lung in time point $i$ is computed as $p_i=\left(r_{i-1,\,i}/u\right)/n$, where $r_{i-1,\,i}$ and $u$ are the number of points on the right lung and the total number of points between time points $i$ and $i-1$, respectively, and $n=20$ is the number of assessed scans.}

\subsubsection{Nodule detection performance}
Similarly to the LUNA16, an annotation is considered as a true-positive ($TP$) if the distance between the ground-truth's and the marking's centroids is less than the nodule's diameter. Also, annotations in non-nodules and multiple hits on the same nodule were neither considered as false-positive or $TP$. Finally, all ground-truth nodules without an annotation were counted as false-negatives ($FN$), and all marks without an associated nodule as false-positive. \hlc{The combination of the annotators is performed via the union of the respective single annotation's sets.} Also, the nodule detection performance of the annotators and the automatic system is evaluated in terms of sensitivity ($TP/(TP+FN)$) \hlc{and average number of false-positive findings per scan.}

The time spent analyzing if an abnormality is indeed a nodule, \hlc{$t_{\, \text{attention}}$} is assessed by summing the values of $\mathcal{A}$ circumscribed by a cylinder of diameter $d$ (5.2cm) and height of the nodule's equivalent diameter centered on its center-of-mass. \hlc{The normalized attention time is defined as $t_{\, \text{attention}}/T$.}    

%The inter-observer agreement of the nodule's characterization is assessed by measuring the absolute distance between two annotations of the same nodule. Similarly, the segmentation-wise agreement between two pixel-wise annotations $s_1$ and $s_2$ is evaluated in terms of the intersection-over-union metric, $IoU= (s_1 \cap s_2) /  (s_1 \cup s_2)$.

\subsubsection{Statistical analysis}

Statistical differences related to detection performance are assessed using an adaptation of the McNemar's test~\cite{McNemar1947NotePercentages}. In this study, this test allows to compare performance of pairs of annotators A and B (including the automatic system) based on their accuracy on an independent test set. Namely, the chi-squared ($\chi^2$) distribution with 1 degree of freedom is defined by Eq.~\ref{eq:mcnemar}:

\begin{equation}
   \chi^2 = \frac{(\,|\,n_{01}-n_{10}\,|-1\,)^2}{n_{01}+n_{10}}
   \label{eq:mcnemar}
\end{equation}

\noindent where $n_{01}$ is the number of nodules not detected by B but detected by A and $n_{10}$ is the number of samples nodules not detected by A but detected by B.
%$\chi^2>3.841$, corresponding to a p-value of 0.05. 
%https://www.mathworks.com/help/stats/one-way-anova.html#bqttdz9-2

Statistical differences related to the elapsed time are assessed via the ANOVA test~\cite{john1996applied}.
Herein, this test is used for assessing if the average elapsed time of analyzing an abnormality or a scan is different between the annotators. The ANOVA test is based on an $F$-distribution with ($k – 1$, $N – k$) degrees of freedom as in Eq.~\ref{eq:anova}: 

\begin{equation}
   F_{k-1,N-k} \sim \frac{SSR}{k-1}\Bigg/\frac{SSE}{N-k}
   \label{eq:anova}
\end{equation}

\noindent where $k$ is the number of annotators, $N$ is the total number of observations, $SSR$ is the variation of the annotator means from the overall mean and $SSE$ is the variation of the observations of each annotator from the respective annotator mean. For both tests, the null hypothesis that the annotators are statistically different is reject if $p$-value$>0.05$.

\section{Results}
\label{sec:results}

\begin{figure}
    \centering
    \includegraphics[width=.8\columnwidth]{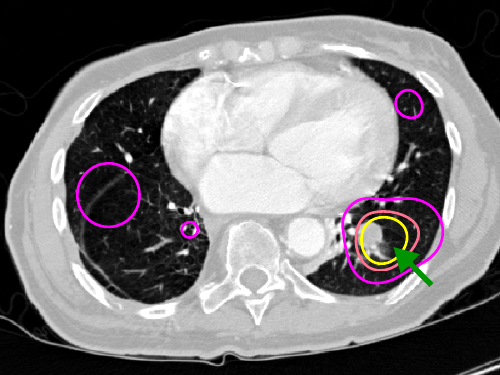}
    \caption{Example of an attention map for an axial slice containing the center of mass of a nodule (arrow). Colorbar: 0~\protect\includegraphics[width=4em,height=.75em]{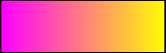}~1%$10^{-4}$ (s)}
    ~(normalized search time in the slice).}
    \label{fig:attention_map}
\end{figure}

\begin{table}[]
    \centering
    \caption{Average normalized attention time (\%) $\pm$ confidence interval ($p=0.05$) used for decisions regarding a potential abnormality. 
    Ann: annotator; TP: true-positive; FP: false-positive; FN: false-negative. \label{tab:nodule_time}}
    \begin{tabular}{|c|c|c|c|} \hline
\textbf{Ann} & \textbf{TP} & \textbf{FP} & \textbf{FN} \\\hline %& \textbf{All} \\\hline
Rad 1 & 10.41$\pm$2.95 & 6.19$\pm$2.79 & 4.45$\pm$1.86 \\\hline
Rad 2 & 10.04$\pm$3.13 & 1.66$\pm$0.39 & 6.24$\pm$4.02 \\\hline
Rad 3 & 14.84$\pm$3.76 & 6.85$\pm$1.53 & 8.34$\pm$6.36 \\\hline
Rad 4 & 10.07$\pm$2.47 & 2.86$\pm$2.29 & 5.35$\pm$3.2 \\\hline
    \end{tabular}
\end{table}

\begin{figure}
    \centering
    \includegraphics[width=\columnwidth]{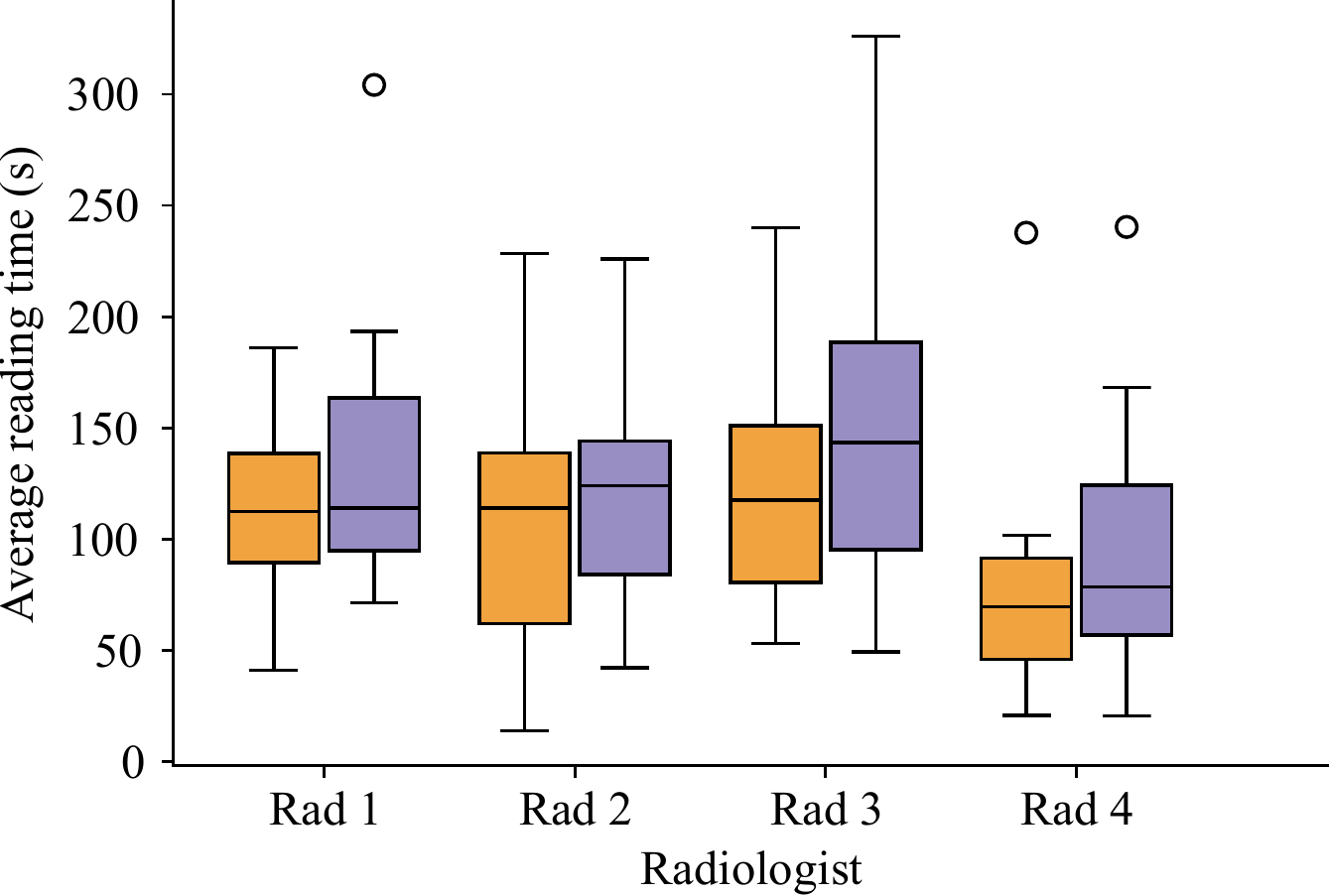}
    \caption{Average reading time for the {\color{light orange}$\blacksquare$}~left and {\color{lavender}$\blacksquare$}~right side of the scan of each reader. Error bar represents the standard deviation. The total reading time of Rad~4 is statistically different from Rad~1~and~3 (significance level $p=0.05$).}
    \label{fig:reading_time}
\end{figure}

%\subsection{Search strategy}

\begin{figure}
    \centering

    \begin{subfigure}[b]{0.23\textwidth}
      \includegraphics[width=\textwidth]{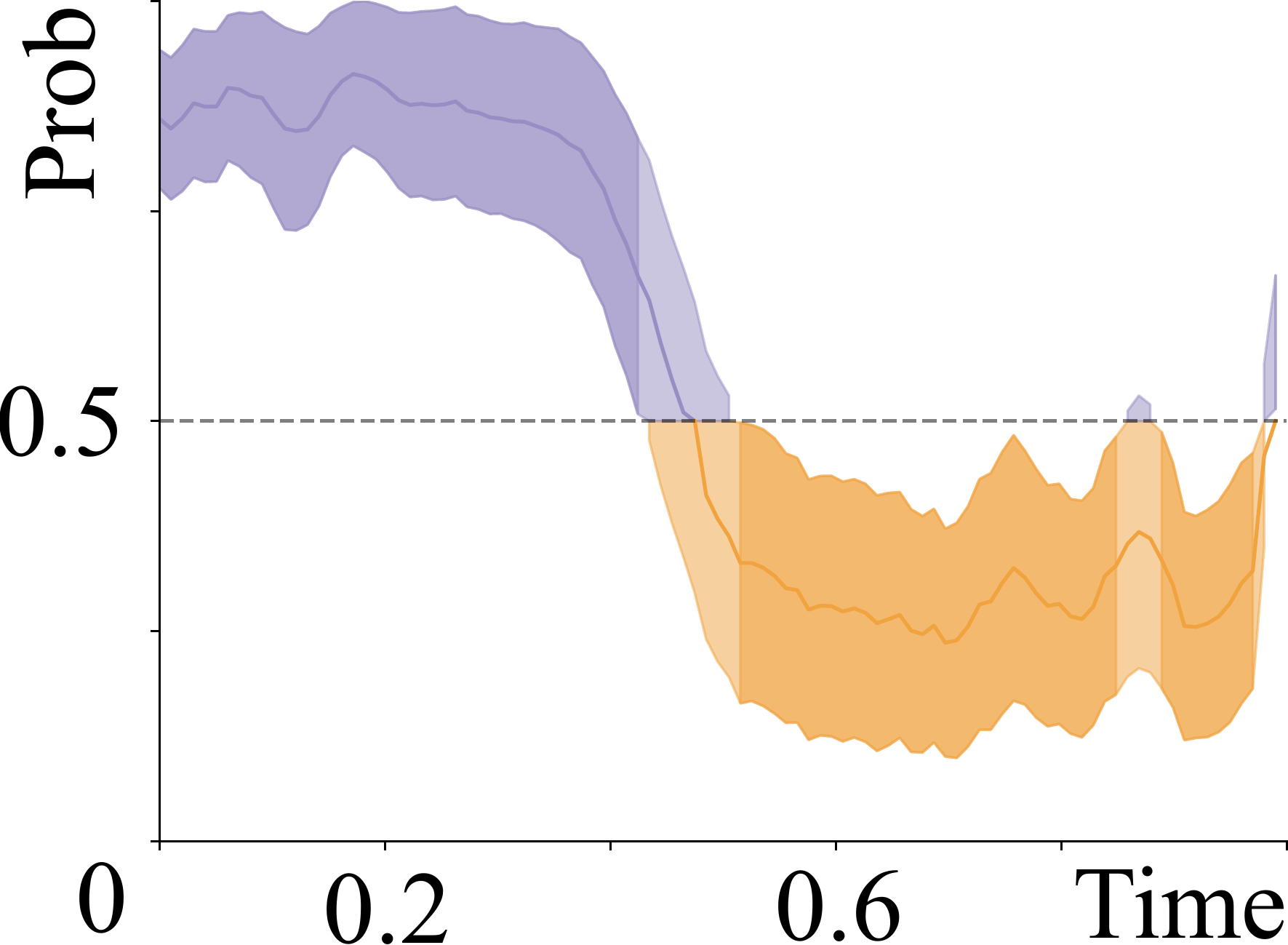}
      \caption{Radiologist 1}
    \end{subfigure}
    \begin{subfigure}[b]{0.23\textwidth}
      \includegraphics[width=\textwidth]{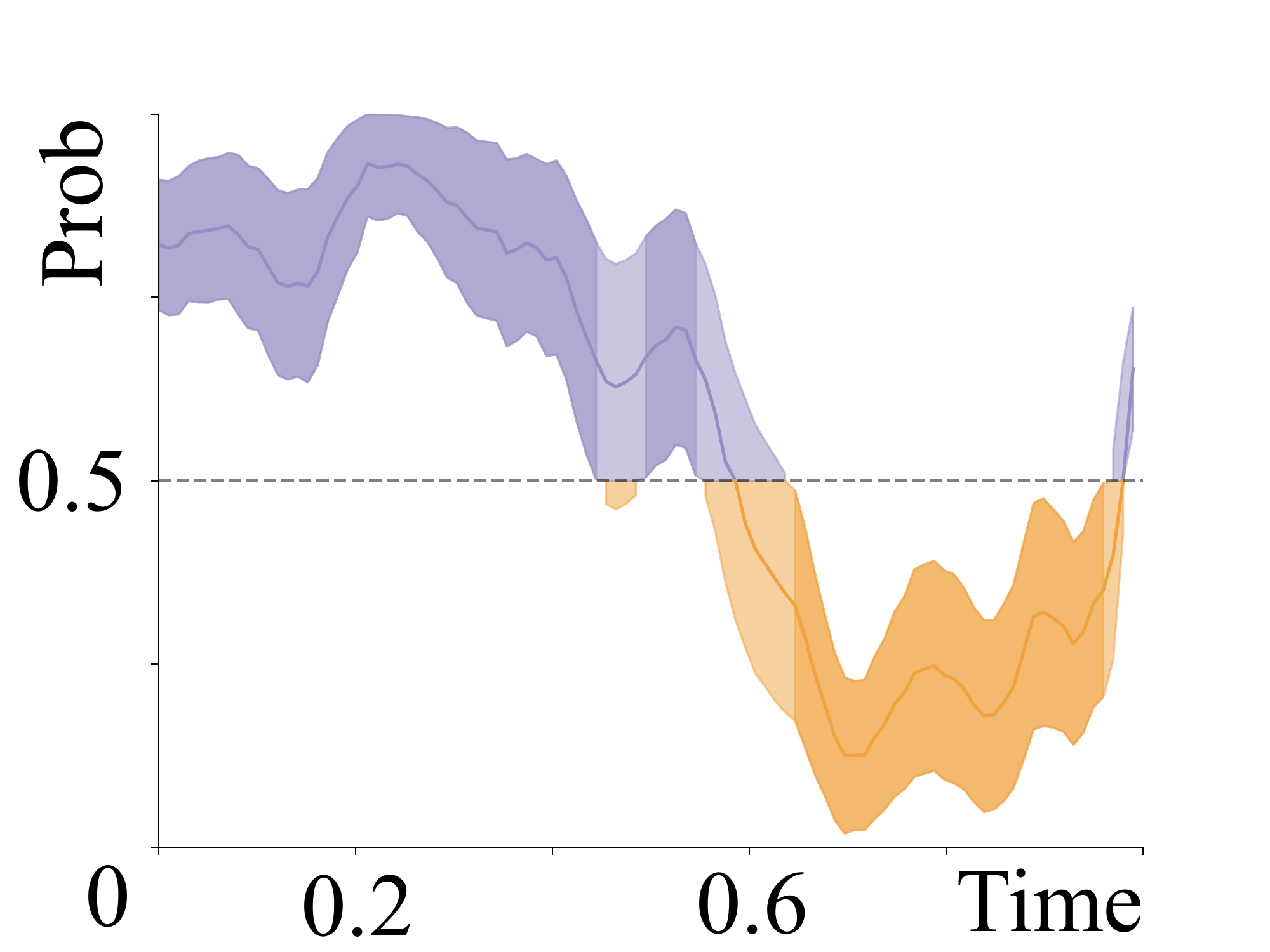}
      \caption{Radiologist 2}
    \end{subfigure}

    \begin{subfigure}[b]{0.23\textwidth}
      \includegraphics[width=\textwidth]{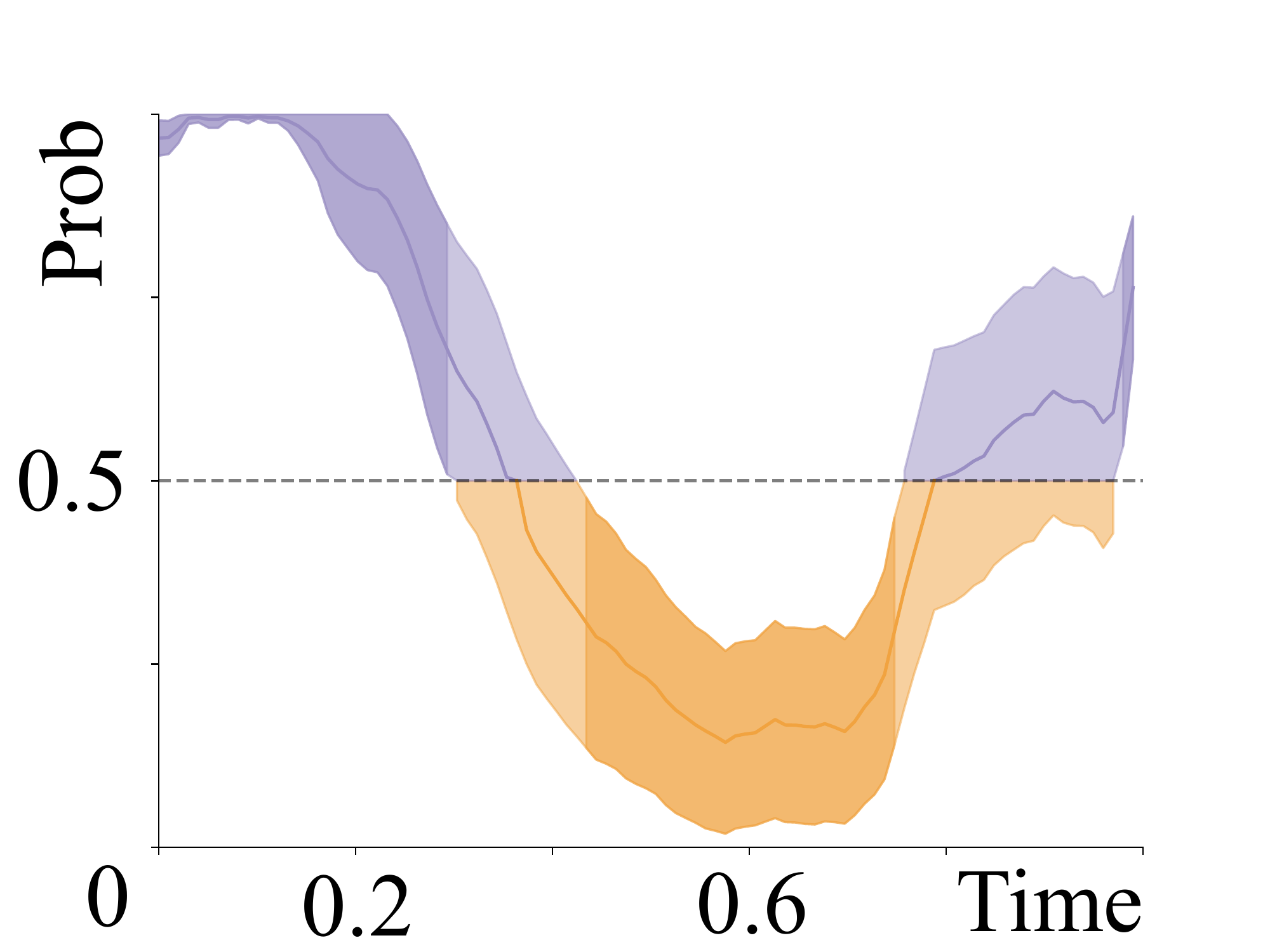}
      \caption{Radiologist 3}
    \end{subfigure}
    \begin{subfigure}[b]{0.23\textwidth}
      \includegraphics[width=\textwidth]{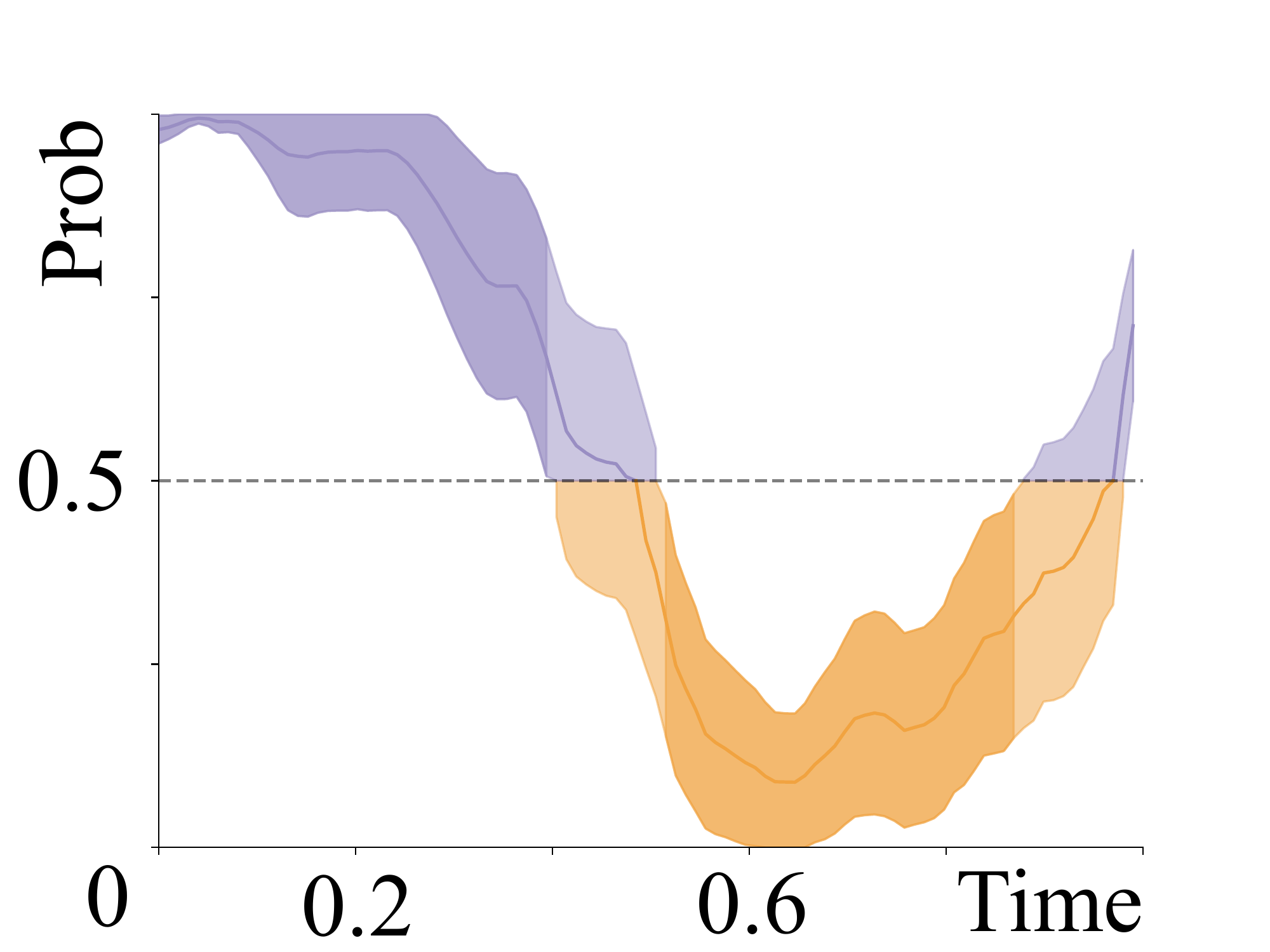}
      \caption{Radiologist 4}
    \end{subfigure}    
    \caption{Probability of the radiologists' gaze location (Prob) for the right lung as function of the normalized scan reading time (Time). Gaze location: {\color{light orange}$\blacksquare$}~left and {\color{lavender}$\blacksquare$}~right sides of the lung. Shaded areas correspond to the confidence interval for $p=0.05$, and darker regions mark periods where the confidence interval is limited to one of the lungs. \label{fig:gaze_strategy}}
    
\end{figure}

\begin{figure*}
    \centering

    \begin{subfigure}[b]{.32\textwidth}
      \includegraphics[width=\textwidth]{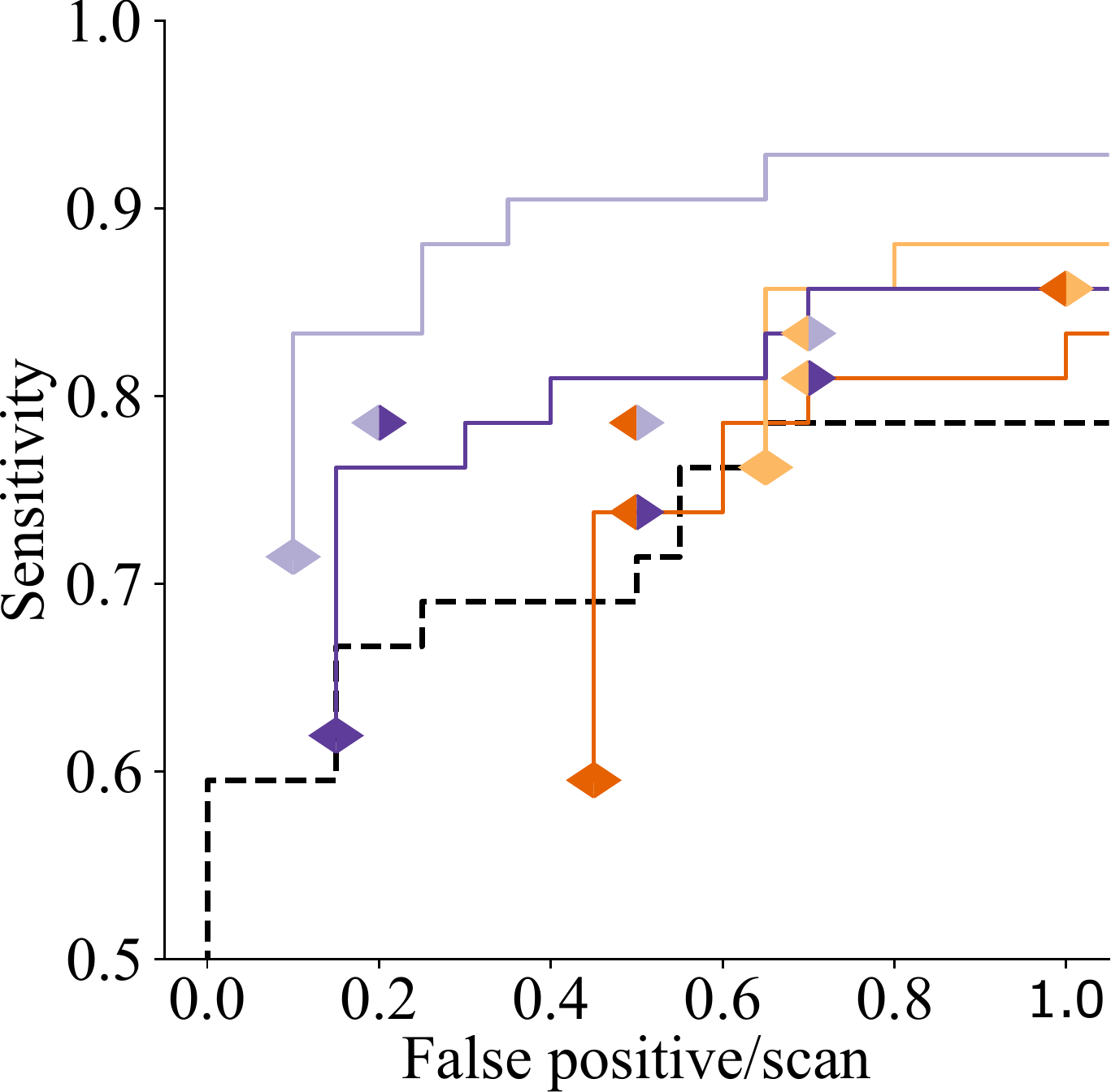}
      \caption{All candidates \label{fig:det_performance_all}}
    \end{subfigure}
    \hfill
    \begin{subfigure}[b]{.32\textwidth}
      \includegraphics[width=\textwidth]{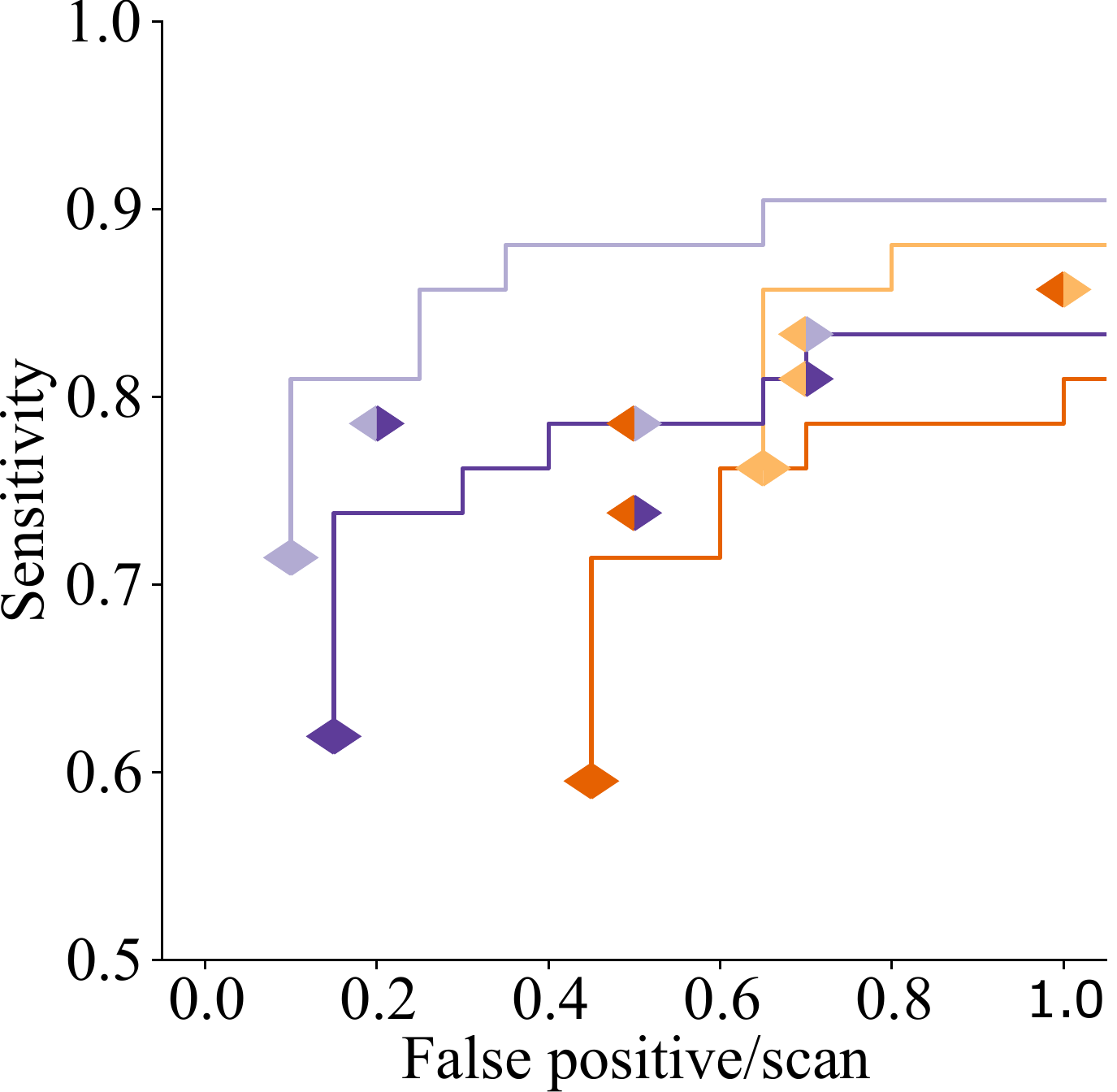}
      \caption{CADe$<$10\% normalized attention time \label{fig:det_performance_10}}
    \end{subfigure}
    \hfill
    \begin{subfigure}[b]{.32\textwidth}
      \includegraphics[width=\textwidth]{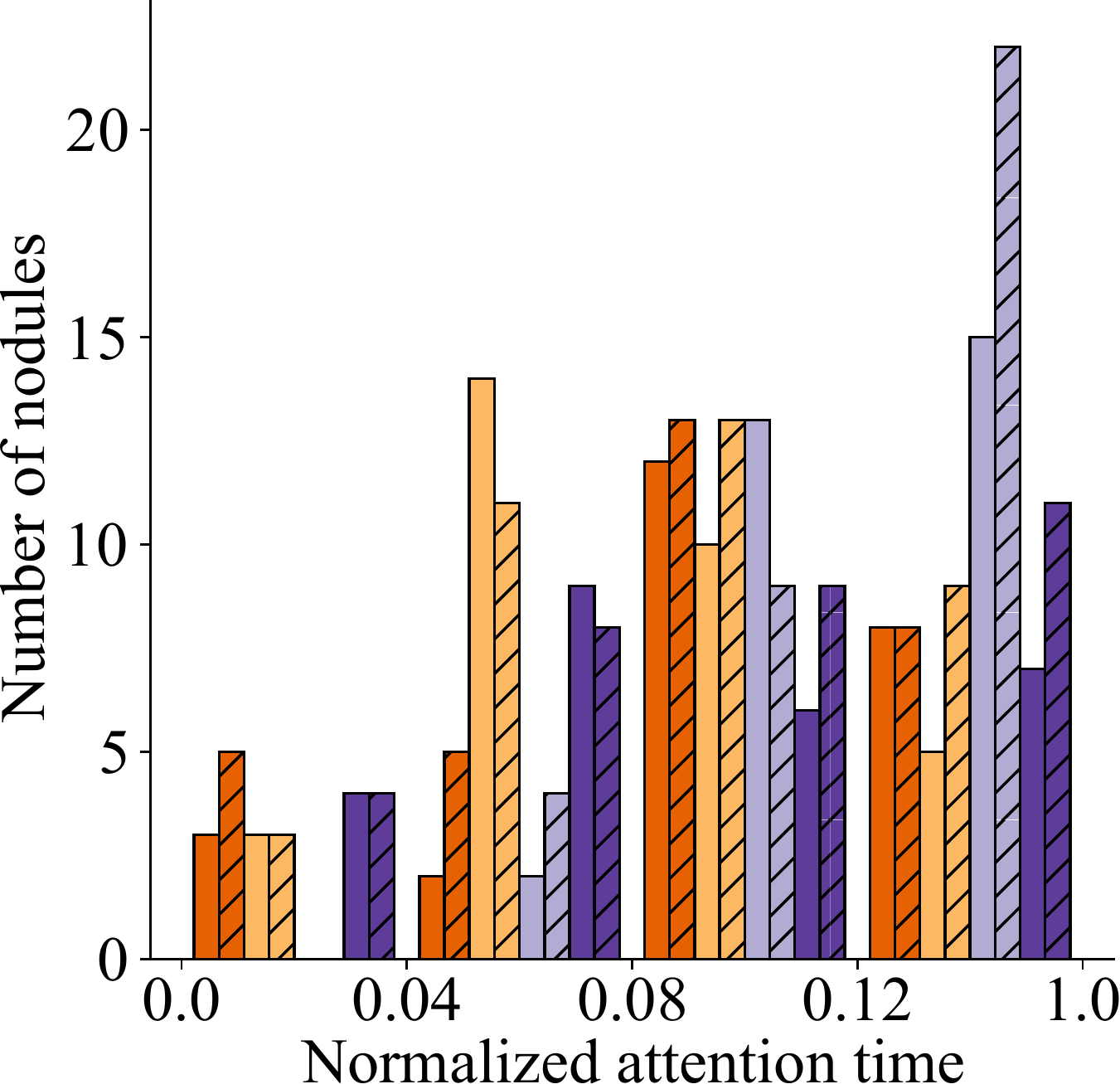}
      \caption{Found nodules \textit{vs} normalized attention \label{fig:det_performance_number}}
    \end{subfigure}

%    \begin{subfigure}[b]{0.32\textwidth}
%      \includegraphics[width=\textwidth]{figures/detection_performance_005.pdf}
%      \caption{Less than 5\% scan reading time}
%    \end{subfigure}
    \caption{Lung nodule detection sensitivity and the corresponding average number of false-positives per scan for the annotators, automatic system and pair-wise combinations. The number of nodules found by ranges of normalized attention time is also shown. 
    {\color{orange}$\blacksquare$}~Rad~1;
    {\color{light orange}$\blacksquare$}~Rad~2;
    {\color{lavender}$\blacksquare$}~Rad~3;
    {\color{purple}$\blacksquare$}~Rad~4;
    {\color{black}$\blacksquare$}~automatic system;
    // combination of the Rad with the automatic system.
      }
    \label{fig:det_performance}
    
\end{figure*}

\subsection{Search technique}

The average reading time for the left and right lungs per observer is shown in Fig.~\ref{fig:reading_time}. In this study, the average scan reading time was $181\pm 84$~(s) and  
the right lung tends to be 20\% more observed than its counterpart. 

The four radiologists use a similar drilling search strategy, as illustrated in Fig.~\ref{fig:gaze_strategy}. 
Specifically, at least 30\% of the initial reading time to assess the right lung ($p=0.05$), then tend to refocus their attention to the left and finally return to the right.

%Namely, there is a tendency to start the scan assessment by the right lung and later the left lung. %Also, the larger uninterrupted periods in which the gaze has a high certainty of being located in just one of the lungs suggests that the radiologists with more experience are more consistent in their search routine. In total, the relative assessment periods with high confidence of the gaze location, \textit{i.e.} where there's a high confidence on the gaze being only on one side of the scan, correspond to .59, .85, .21 and .61 of the entire scans for the radiologists 1, 2, 3 and 4, respectively.

Specialists tend to focus their attention on anatomical feature such as fissures and blood vessels during the nodule hunting, as suggested by Fig.~\ref{fig:attention_map}. Also, Table~\ref{tab:nodule_time} shows that approximately 20\% of the reading time was used for assessing findings that were nodules. Rad~2 was significantly faster than Rad 1~and~3 when marking false-positive findings, but no other statistical differences were found.

%\subsection{Duration of scan assessment}

%Table~\ref{tab:nodule_time} shows that, in average, less than $1\%$ of the reading time is used for deciding if a found abnormality is a nodule. 

\begin{figure*}
    \centering
    \includegraphics[width=.93\textwidth]{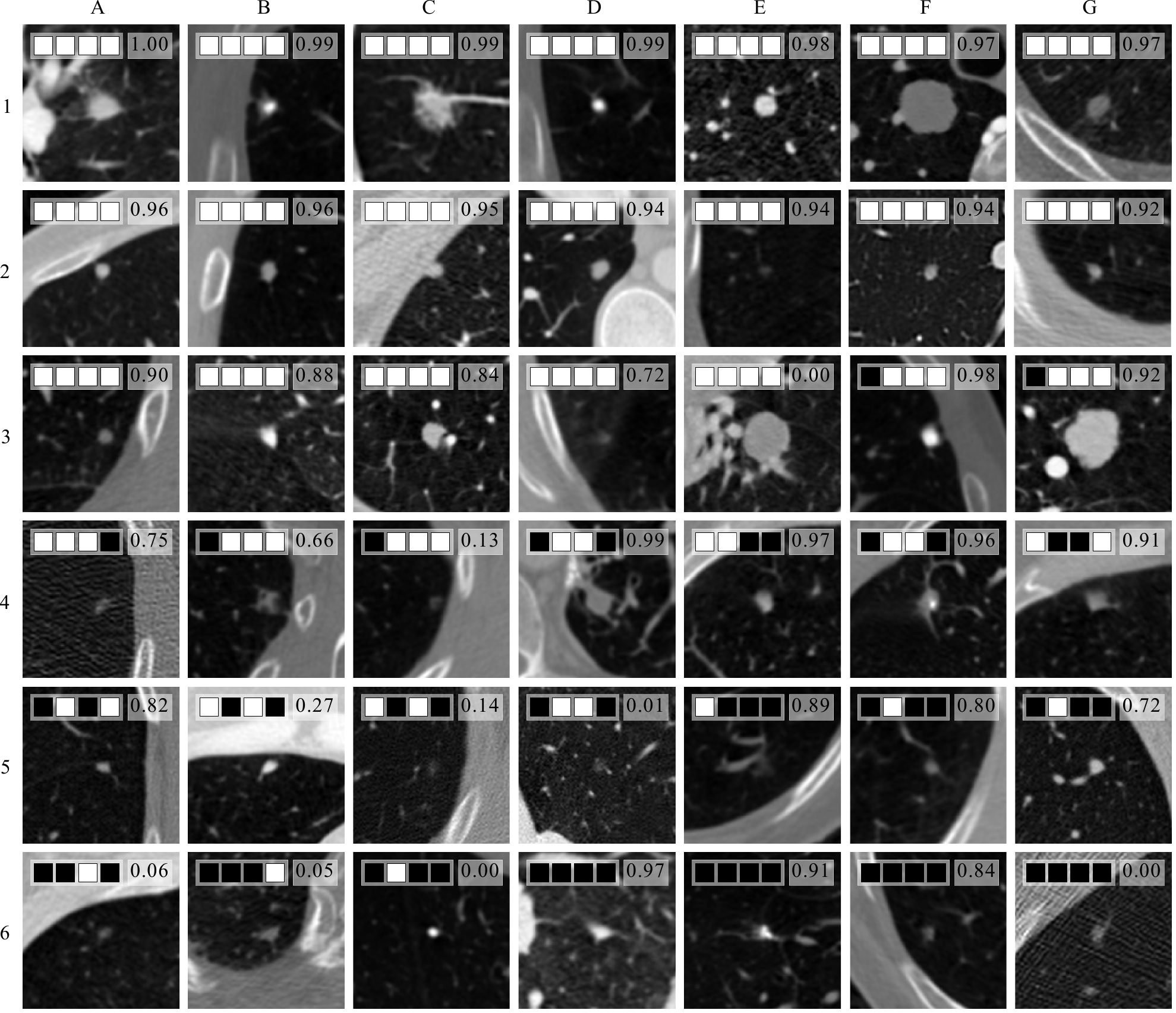}
    \caption{Thumbnails of the central axial view ($64\times64~mm$) of the 42 nodules in this study. {$\blacksquare$}/{$\square$} non-detected/detected by Radiologist 1,2,3~or~4, respectively. The number in the right is the score provided by the automatic system.}
    \label{fig:all_nodules}
\end{figure*}

\subsection{Nodule detection performance}

The overall nodule detection performance of the radiologists and automatic system are depicted in Fig.~\ref{fig:det_performance}~and~\ref{fig:all_nodules}. Specifically, Fig.~\ref{fig:det_performance_all} \hlc{shows the sensitivity of the single and combined annotators for all the studied nodules, whereas} Fig.~\ref{fig:det_performance_10} \hlc{depicts the effect of combining all the annotations of the radiologists with those from the automatic system that had less than 10\% normalized attention time from the corresponding radiologist}. Fig.~\ref{fig:det_performance_number} \hlc{shows the number of found nodules per range of normalized attention time} and   Fig.~\ref{fig:all_nodules}  \hlc{depicts all nodules in the study and the respective detection performance of the specialists and the automatic detection system}. The average human sensitivity is $0.67\pm0.07$ with $0.34\pm0.22$ false-positives/scan. For the same number of false-positives, the automatic system achieves a sensitivity of $0.69$. Sensitivity-wise, \textit{i.e.} ignoring false-positive annotations, all annotators, including the automatic system, are statistical different with exception of Rad~4 in comparison with Rad~1~and~3. Likewise, combining any two annotators statistically increases the detection sensitivity in comparison to the reader alone.

As shown in Fig.~\ref{fig:det_performance}, the \textit{a posteriori} combination of the automatic system allows to increase the sensitivity in average by 0.11 without increasing the number of false-positives. On the other hand, combining two radiologists allows an increase of 0.13 with 76\% more false-positives.
Fig.~\ref{fig:det_performance} also indicates a tendency of increased sensitivity with the time spent assessing an abnormality. Generically, adding the automatic system increases the sensitivity across all relative gaze ranges. For local assessments shorter than 10\% of the reading time, the automatic system still allows to significantly improve the detection performance.

%\begin{figure}
%    \centering
%    \includegraphics[width=\columnwidth]{figures/gaze_sensitivity.pdf}
%    \caption{Nodule detection sensitivity for different relative gazes without and with the combination of the automatic detection {\color{orange}$\blacksquare$}~Rad~1;
%    {\color{light orange}$\blacksquare$}~Rad~2;
%    {\color{lavender}$\blacksquare$}~Rad~3;
%    {\color{purple}$\blacksquare$}~Rad~4; //~combination.
%    }
%    \label{fig:gaze_sens}
%\end{figure}

% single human
% automatic system
% comb human+human
% comb human+system
% comb human+system (<= gaze th)

%\begin{table}[]
%    \caption{\hl{update} Average lung nodule detection performance in terms of sensitivity and false-positive (FP) per scan. %The average number of non-nodule annotations and total annotations per scan is also shown. $\blacksquare$: assessed %annotator(s) (Radiologist 1,2,3,4 and automatic system, respectively).  \label{tab:detec_performance}}
%    \centering
%    \begin{tabular}{|c|c|c|c|c|} \hline
%         Annotator& Sensitivity & FP/scan & Non-nodule/scan & Marks/scan \\ \hline
%        \input{table.tex}
%    \end{tabular}
%\end{table}

%The performance of the automatic lung nodule detection system is not statistically different from any of the human observers ($p=0.05)$ for 0.1 false-positives per scan. 
%Pairing observers tends to increase the detection sensitivity

%average time per scan
%average time per correct annotation
%average time per failed nodule
%average time per incorrect annotation

%average performance alone
%average performance system alone
%combined performance
\section{Discussion}
\label{sec:discussion}

In this study, the radiologists used a drilling strategy to search for abnormalities. Unlike scanning, drilling allows to focus the attention on one region of the lung, reducing the complexity of the search space. Furthermore, drilling allows for higher 3D context, easing the differentiation of nodules and blood vessels. \hlc{By its turn,} the automatic system uses a \hlc{hybrid drilling-scanning} strategy by inferring on stacks of axial slices, \hlc{which also provide 3D context}, and predicting candidates patch-wise \hlc{over the entire scan}.  During the visual assessment of the scans, there was a clear tendency in providing more focus to right lung by 
\begin{inparaenum}[i)]
\item investing more assessing time and
\item starting the reading session with this structure.
\end{inparaenum}
This may be partially related with the left-right top-bottom writing system used on the majority of the occidental countries, since the right lung appears on the left side of the scan and thus is the first on the reading order. However, the most likely explanation is related to medical educational and experience factors. Indeed, it is known that the right lung has a higher probability of containing malignant lesions in comparison with the left lung~\cite{Perandini2016DistributionNature}. Because of this, radiologists may have a tendency to provide more attention to this side of the lung. Further studies should be made to assess the relative influence of these two factors on the search strategy.

The found average human lung nodule detection sensitivity of 0.67 is similar to previous studies~\cite{ArmatoIII200928,AlMohammad2019RadiologistCT}. \hlc{Interestingly, there was no evidence that the average reading time was associated with the detection performance. This is indicative that each specialist has its own reading speed and tend to take a proportional amount of time assessing if found abnormalities.}
\hlc{Indeed, as shown in }Fig.~\ref{fig:all_nodules}, \hlc{all radiologists were capable of locating nodules with a large variety of textures and sizes, including highly subtle abnormalities as E2 and D3. On the other hand, detection failures of non-subtle nodules tended to occur for smaller sizes, as in D6 and E6. The assessment of the individual assessment times indicate that these failures are most likely due to fixation errors. In fact, as shown in} Fig.~\ref{fig:det_performance_number}, \hlc{there is a trend of increasing the number of detection nodules when higher attention times are used.} Interestingly, Table~\ref{tab:nodule_time} suggests that failure, either under- or over-diagnosis is usually associated with lower observation time. \hlc{These findings further indicate the need for automatic second opinion systems, since these can force radiologists to assess unseen abnormalities and possible mitigating attention-related detection failures.} 

The automatic system achieved a nodule detection performance similar to the radiologists, as depicted in Fig.~\ref{fig:det_performance_all}. \hlc{In fact, when combined with human annotations, the system enables a performance increase similar or better than that of two radiologists. These results suggest that the second opinion provided by the automatic system is as valid as a human's, allowing to significantly increase the detection sensitivity without changing the number of false-positive.

Likewise, combining gaze information with the automatic system allows to mitigate failures related to lower observation times. As shown in} Fig.~\ref{fig:det_performance_10}, \hlc{using the CADe system only on regions with less than 10\% normalized attention time (\textit{i.e.} regions where there is a higher false-positive and false-negative probabilities, according to} Table~\ref{tab:nodule_time}) \hlc{ still allows to increase the detection performance. In a scenario where the combination is not done \textit{a posteriori}, as in this study, but instead the radiologist is invited to review the CADe findings after the screening routine, these results suggest that using the gaze-CADe pair could allow to improve the overall detection sensitivity, while reducing the time overhead introduced by the analysis of all CADe findings.}

\section{Conclusion}
\label{sec:conclusion}

Lung nodule hunting is a complex task, but using the opinion of a second radiologist allows to significantly improve the success of the process. This second opinion can be replaced by a properly trained automatic detection system. Also, assessing the gaze during the screening routine allows to retrieve important information related to search strategies and identify potential regions of detection failure. When combining this gaze information with the inferences of an automatic system, it is possible to significantly increase the global detection sensitivity without forcing the radiologist to review the entire volume.
This leads to a less tiresome and faster verification process by reducing the number of candidates to review, while also reducing CADe-related bias since the process is done after the initial assessment. Because of this, the introduction in the clinical practice of systems similar to the one herein presented may contribute to increase the success of lung cancer screening programs by reducing personnel costs and, most importantly, improve the quality of life of the patient.

%\input{conclusion.tex}

\ifCLASSOPTIONcaptionsoff
  \newpage
\fi



\bibliographystyle{IEEEtran}
\bibliography{references.bib}


%\end{IEEEbiography}


%\newpage



%\vfill


\end{document}